\documentclass[11pt,twocolumn]{article}
\usepackage{graphicx}
\usepackage{subfig}
\usepackage{amssymb}
\title{Effect of doping of zinc oxide on the hole mobility of poly(3-hexylthiophene) in hybrid transistors}
\author{$Maria~S.~Hammer^1\footnote{maria.hammer@physik.uni-wuerzburg.de},~Carsten~Deibel^1,~Jens~Pflaum^{1,2}\footnote{jpflaum@physik.uni-wuerzburg.de},~VladimirDyakonov^{1,2}\footnote{vladimir.dyakonov@physik.uni-wuerzburg.de}$\\
\and {\footnotesize  $^1$~Experimental~Physics~VI,~Faculty~of~Physics,~Julius~Maximilians~University~of~W\"urzburg, Germany}\\
\and {\footnotesize $^2$Functional~Materials~for~Energy~Technology,~Bavarian~Centre~for~Applied~Energy~Research~(ZAE~Bayern), Germany}\\
\\
\and accepted~for~pulication~in~Organic~Electronics}
\begin{document}
\date{}
\maketitle

\begin{abstract}
Hybrid field effect transistors based on the organic polymer poly(3-hexyl\-thio\-phene) (P3HT) and inorganic zinc oxide were investigated. In this report we present one of the first studies on hybrid transistors employing one $polymeric$ transport layer. The sol--gel processed ZnO was modified via Al doping between 0.8 and 10~at.$\%$, which allows a systematic variation of the zinc oxide properties, i.e. electron mobility and morphology. With increasing doping level we observe on the one hand a decrease of the electron mobilities by two orders of magnitude,%from 3.2$\cdot$10$^{-3}$~cm$^2$V$^{-1}$s$^{-1}$ to 6.5$\cdot$10$^{-5}$~cm$^2$V$^{-1}$s$^{-1}$, 
on the other hand doping enforces a morphological change of the zinc oxide layer which enables the infiltration of P3HT into the inorganic matrix. X-ray reflectivity (XRR) measurements confirm this significant change in the interface morphology for the various doping levels. We demonstrate that doping of ZnO is a tool to adjust the charge transport in ZnO/P3HT hybrids, using one single injecting metal (Au bottom contact) on a SiO$_2$ dielectric. We observe an influence of the zinc oxide layer on the hole mobility in P3HT which we can modify via Al doping of ZnO. Hence, maximum hole mobility of 5$\cdot$10$^{-4}$~cm$^2$V$^{-1}$s$^{-1}$ in the hybrid system with 2~$\%$ Al doping. 5~at.$\%$ Al doping leads to a balanced mobility in the organic/inorganic hybrid system but also to a small on/off ratio due to high off-currents. 

\end{abstract}

\section{Introduction}

The fabrication of efficient printable electronics demands the development of solution processed complementary circuits, where both n-type and p-type semiconductors are incorporated using a simple device architecture. In this context, ambipolar transistors employing a single organic compound as active layer have been reported~\cite{schmechel2005}. A drawback of the latter device is, though, that the electron and hole mobility often differ and, technologically more demanding, two different metals have to be used for electron and hole injection. It has also been shown that by simply blending organic p- and  n-conductors the desired ambipolarity can be achieved in transistors, e.g. using [6,6]-phenyl C61-butyric acid methyl ester (PCBM) and poly[2-methoxy-5-(3$'$,7$'$-dimethyloctyloxy)]-p-phenylene vinylene~(OC$_1$C$_{10}$-PPV)~\cite{meijer2003}. However, a non--Ohmic contact behaviour for electron injection from Au into PCBM has been observed here. 
A promising approach presented in this paper is the use of an organic/inorganic hybrid systems. Inorganic materials offer the customizability of the charge transport properties via doping and therefore an optimum ambipolar performance in combination with the organic semiconductor. The organic counterpart offers all advantages of polymer technologies, e.g. printability and infiltration ability. Furthermore, doping results in structural modifications of the inorganic component, such as surface roughness and porosity, which allows to define a three dimensional matrix for the subsequent intrusion of the organic polymer. 
Though inorganic nanocrystalline zinc oxide has been successfully used in hybrid solar cells~\cite{beek2005} and bilayers of the thermally evaporated pentacene on ZnO have been investigated~\cite{pal2008}, the advantages of ZnO--polymer hybrids for ambipolar field effect transistors have not been explored yet. 

In this work we used sol--gel processed ZnO nanoparticles with various Al doping levels, as the n-type semiconductor, and regioregular poly(3-hexyl\-thio\-phene) (P3HT), as the p-type component, to fabricate field effect transistors (FETs) with gold source and drain contacts on a SiO$_2$ dielectric. 
Our previous work~\cite{hammer2008} evinced, $inter alia$, that doping has a strong effect on the structural and electrical properties of sol--gel processed sintered ZnO:Al. The electron mobility is slightly decreasing due to structural, morphological, and ionic defects. 

For the fabrication of the hybrid system presented in this paper, it was necessary to omit the sintering step for the Al doped ZnO in order to allow intrusion of P3HT into the ZnO-matrix. 
We show that by doping the zinc oxide nanoparticles as well as by morphological changes in the blend it is possible to adjust a balanced electron and hole mobility in the hybrid material, which is indispensable for a good performance of an ambipolar transistor. The presented data will be discussed in terms of the charge transport mechanisms in field effect transistors together with their morphology of the thin films which was monitored via X-ray reflectivity (XRR) measurements.

\section{Experimental}
\subsection{Materials and Synthesis}
The synthesis of Al doped ZnO was performed according to the route introduced by Xue~$et$~$al.$~\cite{xue2006}.
2-methoxyethanole (MTE), monoethanolamine (MEA), zincacetate dihydrate (ZAD) and aluminiumnitrate nonahydrate (ANN) were purchased from Sigma-Aldrich. 2.1 ml of the stabilizer MEA were dissolved in 100~ml of the solvent MTE to lead to a 0.35~M solution. 35~mmol of the starting material ZAD were added. Depending on the designated doping level ANN was admixed to the solution to achieve up to 10~at.$\%$ Al doping. The solution was stirred for 2~hrs. at 60$^{\circ}$\space C and subsequently aged for at least one day.

Regioregular poly(3-hexyl\-thiophene) (P3HT) was purchased from Rieke Metals and used without further purification. P3HT was dissolved in chlorobenzene at 0.1~wt$\%$.

\subsection{Sample Preparation and Measurement}
The field effect transistor structures were prepared on Sb-doped silicon wafers purchased at Si-Mat. The highly n-doped Si (001) with a nominal resistivity of about 0.02~$\Omega$m serves as the gate electrode. The top layer of the wafer consists of 200~nm thermally grown SiO$_2$ which forms the dielectric. Photolithography was used to get well defined source--drain structures. 1~nm Ti as an adhesion layer and subsequently 19~nm thick Au contacts were thermally evaporated on top of the dielectric. After cleaning the substrate with acetone and isopropanol, the ZnO solution with various Al doping level was spin coated for one minute at 3000~rpm. Thereafter, the substrate was heated to 300$^{\circ}$\space{}C in an oven in ambient atmosphere for 100 minutes in order to remove MEA. Then the samples were transferred to a glovebox where the P3HT solution was spin coated. Finally, the samples were slowly dried for 15~hrs.
%The electronic reproducibility of FETs was confirmed for several samples. The relative error was smaller than 5~$\%$ and 10~$\%$, for the bare ZnO:Al and ZnO:Al/P3HT-transistors, respectively. The estimates for the conductivity and the charge carrier density show relative errors of about 20~$\%$

The current--voltage measurements were carried-out using an Agilent Parameter Analyzer 4155C. The room temperature measurements were performed in nitrogen atmosphere before and after the infiltration of P3HT.
In either transistor measurement we chose channel lengths of 160~$\mu$m, so that the channel resistance dominates the contact resistance and, hence, minimizes the influence of the contact on the measured transport parameters~\cite{hill2005}. 

The X-ray reflectivity measurements were carried out using a Diffractometer General Electrics 3003 T-T in Bragg-Brentano geometry. The X-ray measurement provides a non-destructive way of probing the layer thickness, interface roughness and the electron density of single and multi layers. 
For XRR, the sample is being illuminated with Cu~K$_\alpha$ radiation at the characteristic wavelength of $\lambda$=1.5406~\AA. Typically, the reflected intensity is being plotted $vs.$ the wave vector transfer $q_z$ perpendicular to the surface.
$q_z$ and $\theta$ are related via
$q_z=4 \pi/\lambda \sin \theta$.
From $\theta~=~$0$^{\circ}$\space up to a critical angle $\theta_c$, total reflection dominates, as radiation can penetrate the layer only evanescently. By the relation $\theta\sim\sqrt{\rho}$, $\theta_c$ is a measure for the electron density $\rho$ and hence, a measure for the effective material density.
The intermixing of the two phases can be determined from the attenuation of the corresponding Kiessig fringes.

\section{Results}
\subsection{Field effect transistor measurements} \label{fet0at}

The hybrid FETs were built upon a device structure using a SiO$_2$ dielectric and Au source and drain bottom contacts in two steps, first spin coating and heating the inorganic component, subsequently deposition of the organic component. This approach resulted in Ohmic--like contacts for electron and hole injection. The electrical parameters of the devices were investigated before and after the deposition of P3HT.

In a field effect transistor charge carriers are accumulated via a source--gate voltage $V_g$ where the hole or electron accumulation are determined by the sign of $V_g$, corrected for the threshold voltage $V_{th}$. This provides a way to selectively determine the electron and hole mobility at various carrier densities from the transfer characteristics in the linear and saturation regime, $\mu_{lin}$ and $\mu_{sat}$, respectively, via 
\begin{equation}
\mu_{lin}=\frac{L}{wC_i V_d}\frac{\partial I_d}{\partial V_g}\;\;\mathrm{and}\;\;
\mu_{sat}=\frac{2L}{w C_i}\left(\frac{\partial\sqrt{I_d}}{\partial V_g}\right)^2,
\label{mu}
\end{equation}
where $L$ and $w$ are the channel length and width, $C_i$ is the dielectric capacitance per unit area, $I_d$ is the current between source and drain and $V_d$ is the source--drain field.

\begin{figure} 
   \centering
   \includegraphics[width=8cm]{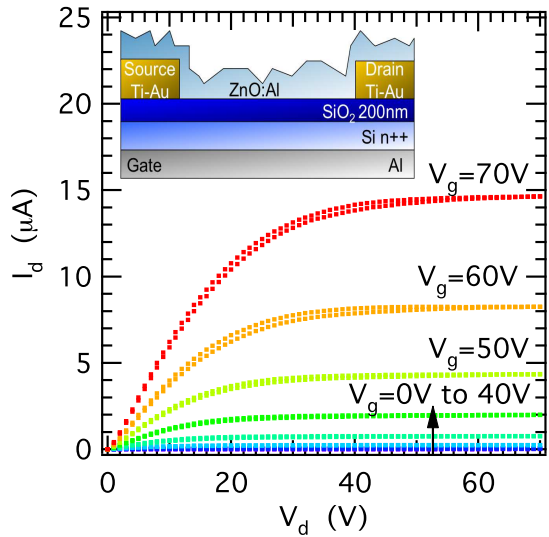} 
   \caption{(Color online) ZnO:Al (2~at.$\%$) output characteristic. Transistor details: Au bottom contacts $L$=160~$\mu$m, $w$=62.4~mm, SiO$_2$ thickness $d$=200~nm. The inset shows a schematic drawing of the device.}
   \label{fig:znopureout}
\end{figure}

\begin{figure} 
   \centering
   \includegraphics[width=8cm]{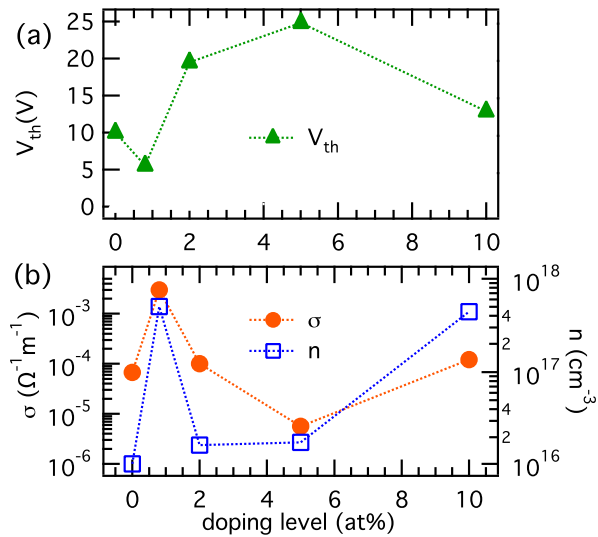} 
   \caption{(Color online) Bare ZnO:Al: (a) Threshold voltage for electron transport in the saturation regime. (b) Estimated conductivity and free charge carrier density. As indicated, there is a defined correlation between $V_{th}$ and $n$.}
   \label{fig:sigmanVt}
\end{figure}

An output characteristic of a transistor with unsintered 2~at.$\%$ Al doped ZnO as an active layer is shown in Fig.~\ref{fig:znopureout}. The gate and the drain voltage, respectively, have been switched between 0~V and 70~V. The output characteristic exhibits Ohmic behavior in the linear regime, indicating efficient injection from the gold contact into the conduction band of ZnO:Al and saturation at high drain voltages. The pinch-off voltage increases quadratically with the applied gate voltage. The threshold voltage $V_{th}$ (see Fig.~\ref{fig:sigmanVt}) has been extracted from the transfer characteristic in the saturation regime. $V_{th}$ is a measure for trapped charges in the material and at the ZnO:Al/dielectric interface and also for free charge carriers intrinsically present in the channel. It has to be noted that additional charge carriers introduced via Al doping will compensate these electron traps. The threshold voltage of ZnO:Al transistors varies between 5.5~V for the 0.8~at.$\%$ Al doped transistor and 25~V for the 5~at.$\%$ doping (see Fig.~\ref{fig:sigmanVt}a).

In order to investigate the electrical properties of ZnO:Al, we estimated the conductivity $\sigma$ in ZnO:Al from the linear regime (low $V_d$) at $V_g$=0~V using $I_d=jwd_{Au}=\sigma V_d/L\cdot wd_{Au}$, where $d_{Au}$ is the thickness of the gold contact, $w\cdot d_{Au}$ is the area of the injecting contact (see Fig.~\ref{fig:sigmanVt}b). We observe a maximum in the estimated effective conductivity of 3$\cdot$10$^{-3}~\Omega^{-1}$m$^{-1}$ for the 0.8~at.$\%$ Al doped ZnO layer, as expected for ZnO:Al~\cite{hammer2008}: Doping presumably leads to higher charge carrier densities but also leads to additional scattering centers. With $\sigma=\mu n e$, where $\mu$ is the mobility at low gate and drain voltages and $e$ is the elementary charge, we are able to provide a rough estimate of the charge carrier density $n$. The charge carrier density varies between 1$\cdot$10$^{16}$~cm$^{-3}$ in the nominally undoped case and 5$\cdot$10$^{17}$~cm$^{-3}$ for the 0.8~at.$\%$ Al doped layer. Additionally, we observe a relatively high charge carrier density for the highest doping level of 10~at.$\%$. The threshold voltage $V_{th}$ reflects the density of free charge carriers, i.e. the high $n$ value for 0.8~at.$\%$ Al doping corresponds to a low $V_{th}$ of 5.5~V.

\subsubsection*{Hybrid ZnO:Al/P3HT-Transistor}
\begin{figure} 
   \centering
   \subfloat[]{\includegraphics[width=8cm]{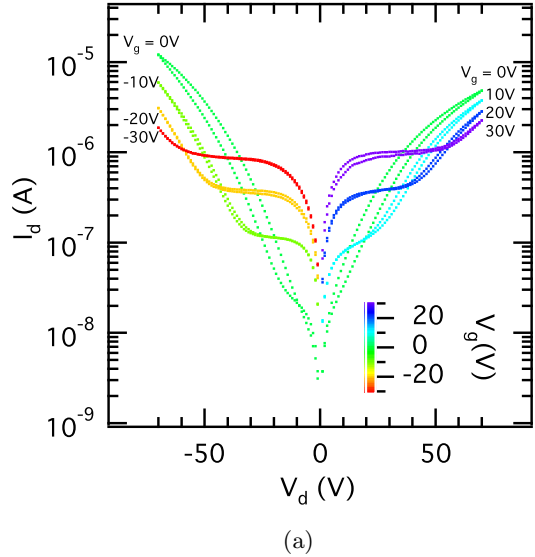}\label{au-p200}}%
	\qquad
	\subfloat[]{\includegraphics[width=8cm]{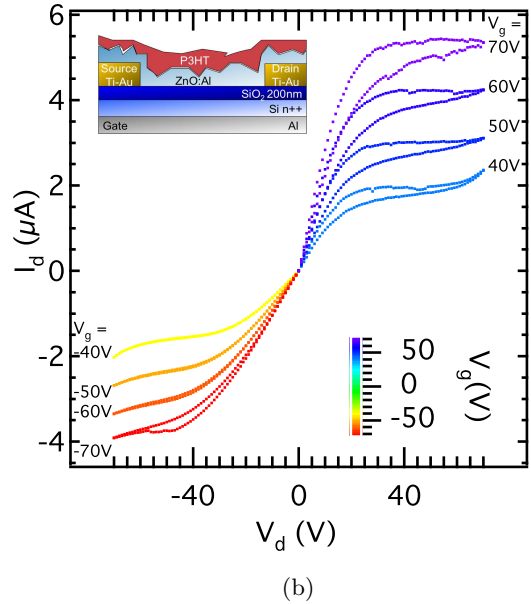}\label{au-p1200}}%
   \caption{(Color online) ZnO:Al(2~at.$\%$)/P3HT-transistor: Output characteristic; $V_g$ is indicated by the color scale and by the legend. Shown is the same device as in Fig~\ref{fig:znopureout} but with additional P3HT hole transport layer: (a) from $V_g$=-30V to 30V, (b) from $V_g$=-70~V to -40~V and from 40~V to 70~V. The inset shows a schematic drawing of the device.}
   \label{fig:2atout}
\end{figure}

After the characterization of the field effect transistors with bare ZnO:Al at several doping levels, the the electronic behaviour of transistors with ZnO:Al/P3HT layers were analyzed..
An output characteristic of the 2~at.$\%$ Al doped ZnO/P3HT transistor is shown in Fig.~\ref{fig:2atout}. 
At high $V_d$ and low $V_g$ both, electrons and holes can be injected into the channel and form a pn-junction which leads to a diode-like behaviour. Applying high negative (positive) $V_g$ and low $V_d$ depletes the electrons (holes) from the channel and the device behaves as an unipolar hole (electron) accumulating transistor. Furthermore, the unipolar regime in both, hole and electron accumulation regions, shows Ohmic behaviour at low $V_d$. It has to be noted that we reached this Ohmic behaviour using gold for electron $and$ hole injection and no additional modification of the dielectric was applied, making the device architecture straightforward and simple to prepare. Comparing the output characteristic of the bare ZnO:Al(2~at.$\%$) FET (see Fig.~\ref{fig:znopureout}) to the device, where P3HT had been coated, we observe only a slight decrease in the maximum drain current. Also, in the electron accumulation regime, an enhanced hystereses can be observed, which indicates additional trap states. We note that with Au contacts no hole accumulation could be observed in bare ZnO:Al.

\begin{figure} 
   \centering
   \includegraphics[width=8cm]{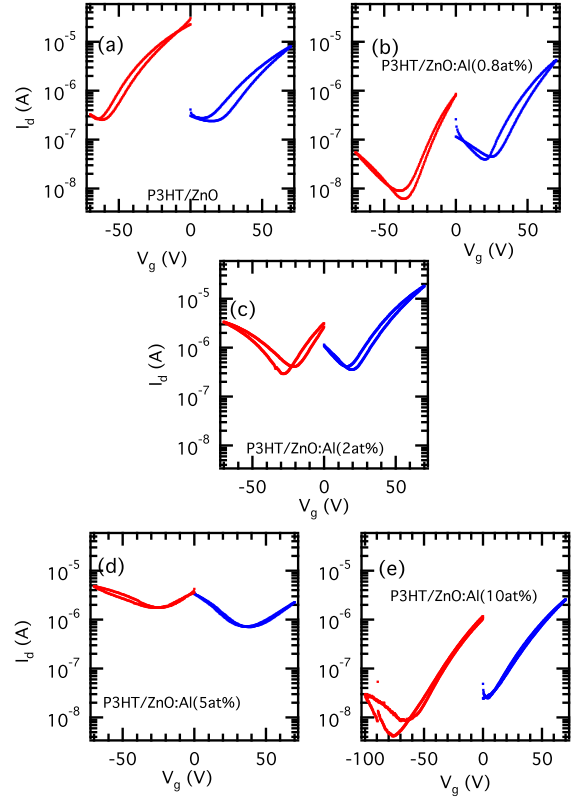} 
   \caption{(Color online) ZnO:Al/P3HT-Transistor: (a)-(e)Transfer characteristics in saturation regime at $\left|V_d\right|$=40~V for hole (red line) and electron (blue line) acumulation. Forward and backward direction show little charging effects. V$_g$ ranges from -70~V to +70~V, except for the 10~at$\%$ system (e).}
   \label{fig:transfersat}
\end{figure}

The transfer characteristics of the P3HT/ZnO:Al transistors at various doping level in saturation at $\left| V_d \right|$=40~V are shown in Fig. \ref{fig:transfersat}(a)-(e). 
The undoped system exhibits unipolar behaviour for electrons. Similar values for the current at high positive $V_g$ can be observed at negative $V_d$ and low gate voltages although hole accumulation occurs only below $V_g=-$50~V. We attribute the high current at $V_g=$0~V and $V_d=-$40~V to an enhanced efficiency of electron $and$ hole injection upon applying a negative bias. The difference between the drain current at $V_g$=0~V at positive and negative $V_d$ is directly correlated to the difference in the hole and electron mobility, which will be discussed in the following. 
Almost symmetric behaviour can be observed for the 2~at.$\%$ doped system.
The transfer characteristics of the 5~at.$\%$ doped ZnO/P3HT has the drawback of a small on/off ratio. For 10~at.$\%$ doping, the onset of hole accumulation occurs at negative voltages higher than $\left| V_g \right|$=70~V.

\begin{figure} 
   \centering
   \includegraphics[width=8cm]{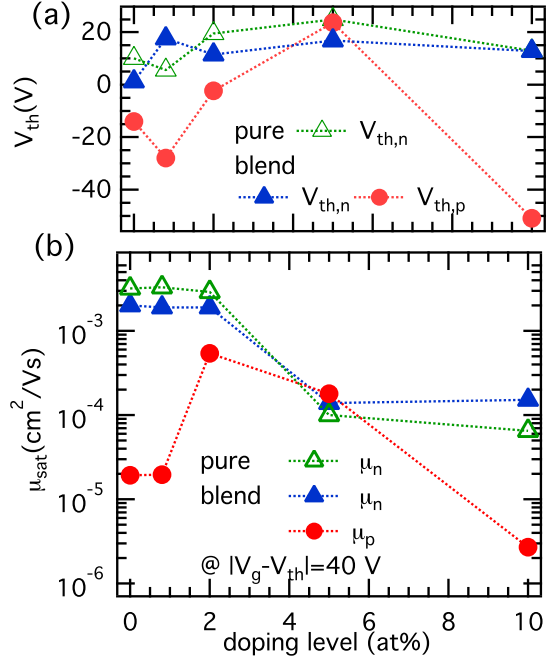} 
   \caption{(Color online) ZnO:Al/P3HT-transistor: (a) Threshold voltage V$_{th,n}$ for electrons and V$_{th,p}$ for holes at various doping levels. V$_{th,n}$ for bare ZnO:Al has been included for comparability, same as in Fig.~\ref{fig:sigmanVt}} (b)Saturation mobility derived from the transfer characteristics; green triangles: electron mobility in bare ZnO:Al as reference, blue full triangles and red circles: electron and hole mobility in ZnO:Al/P3HT, respectively.
   \label{fig:vt_mun}
\end{figure}

Fig.~\ref{fig:vt_mun}(a) shows the threshold voltage $V_{th}$ $vs.$ doping level for electrons ($V_{th,n}$, closed symbols) and holes ($V_{th,p}$, open symbols) for the saturation regime. Whereas $V_{th}$ for the electron operating mode increases from around 0~V in the undoped case to to 17.5~V and stays nearly constant for higher doping levels, we observe large fluctuations of values for the threshold voltage of holes between $+$23.6~V and $-$50.9~V. 

The electron mobility, $\mu_e$, at $\left|V_g-V_{th}\right|$=40~V and $\left|V_d\right|$=40~V is shown in Fig. \ref{fig:vt_mun}(b).
$\mu_e$ decreases over one order of magnitude from 3.2$\cdot$10$^{-3}$~cm$^2$V$^{-1}$s$^{-1}$ for the undoped system to 6.5$\cdot$10$^{-5}$~cm$^2$V$^{-1}$s$^{-1}$ at an Al concentration of 10~at.$\%$. This behaviour is attributed to additional defects of structural, ionic and/or electronic origin introduced via doping~\cite{hammer2008}. Obviously, the infiltration of P3HT does not significantly affect the electron mobility, 
$\mu_e$ in the hybrid system decreases from 2$\cdot$10$^{-3}$~cm$^2$V$^{-1}$s$^{-1}$ to 2$\cdot$10$^{-4}$~cm$^2$V$^{-1}$s$^{-1}$.

The hole mobility (see Fig. \ref{fig:vt_mun}(b)) in the hybrid system, without Al doping and at a doping level of 0.8~at.$\%$, is about two orders of magnitude lower than the hole mobility observed in bare P3HT transistors (not shown). 2~at.$\%$ Al doping of ZnO induces an increase of the hole mobility of more than one order of magnitude to 5$\cdot$10$^{-4}$~cm$^2$V$^{-1}$s$^{-1}$.

In the case of hole accumulation, ZnO:Al can be assumed to be an additional dielectric.
The accumulated charge per unit area $Q$ depends on the thickness of the dielectric 
\begin{equation}
Q=\epsilon \epsilon_0/d \cdot \left(V_g-V_{th}\right), 
\label{q}
\end{equation}
where $\epsilon$ and $\epsilon_0$ are the relative and absolute permittivity of the dielectric and $d$ is the thickness of the dielectric. Note that the calculated hole mobilities refer to a 200~nm SiO$_2$ dielectric solely. The assumption of a constant capacitance was verified experimentally. Capacitance measurements on Si~n$^{++}$/SiO$_2$(200~nm)/ZnO:Al/Au, i.e. treating SiO$_2$ and ZnO:Al as two series capacitors did not reveal significant changes with respect to the bare Si$O_2$ dielectric. This result is plausible since the smallest capacitance, here that of SiO$_2$, determines the total serial capacitance. 

Evidently, the undoped and the 0.8~at.$\%$~Al doped ZnO/P3HT hybrid transistors do not exhibit balanced electron and hole mobilities. $\mu_e$ and $\mu_p$ differ by two orders of magnitude. In the case of 2~at.$\%$ Al doping we observe a hole mobility of 5$\cdot$10$^{-4}$~cm$^2$V$^{-1}$s$^{-1}$ and the value differs by only a factor of five from $\mu_e$ at the same doping level. For a doping levels of 5~at.$\%$ we indeed obtain balanced mobilities and aside from that, a decrease of the electron mobility. The highest Al doping level of 10~at.$\%$ leads to a strong decrease in $\mu_p$ and a large difference between $\mu_p$ and $\mu_e$.

\subsection{Interface morphology}
The dependence of the P3HT hole mobility on the Al doping level of ZnO cannot be understood merely from the data of the electrical measurements. The relation between $\mu_p$ and doping level could have several reasons: 

(i) the crystallinity of P3HT, 

(ii) the interface condition of ZnO and the related ability to infiltrate P3HT, 

(iii) the dielectric properties at the interface

(iv) defects or charged traps at the organic/inorganic interfaces generated by doping.

Obviously, in any of the four cases the interface morphology is of essential relevance for the density of free holes in the FET channel.

\subsubsection*{X-ray reflectivity}
The intermixing of both semiconductor phases at the interface can be quantified via X-ray reflectivity measurements: XRR is sensitive to the electron density of the materials which differs by one order of magnitude for P3HT and undoped/Al doped ZnO. XRR has the unique advantage over, e.g. AFM, that buried interfaces can be probed on the complete sample.

\begin{figure} 
   \centering
   \includegraphics[width=8cm]{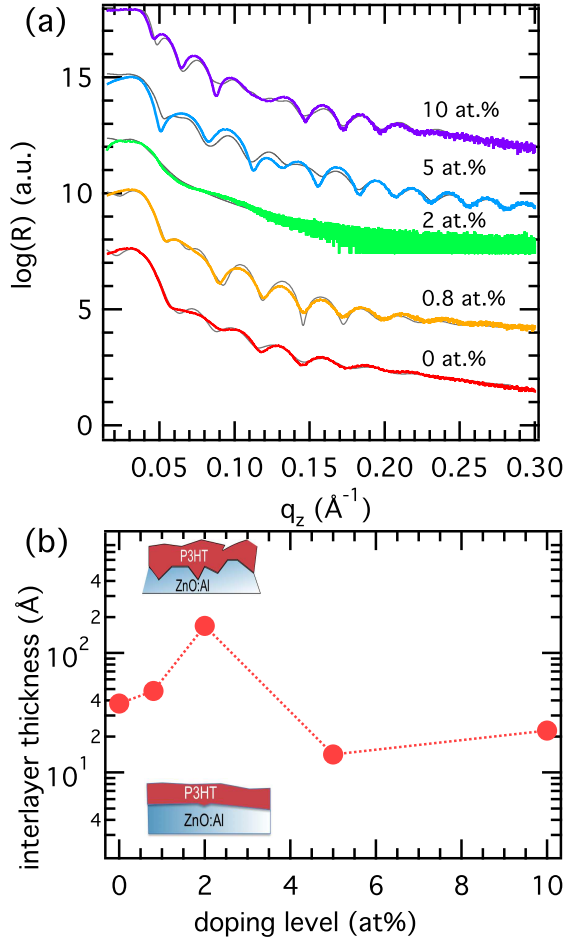} 
   \caption{(Color online) (a)  X-ray reflectivity curve of nc-ZnO:Al/P3HT at various Al doping levels as indicated by the legend. Grey lines are corresponding to fits based on the Paratt formalism. The profiles have been offset for clarity. (b) ZnO:Al/P3HT effective layer thickness of intermixing as derived from XRR spectra. A schematical drawing of intermixing is provided by the insets: upper drawing: high level of intermixing, lower drawing: low level of intermixing.}
   \label{fig:xrr}
\end{figure}

The XRR measurements of ZnO/P3HT for different Al doping levels are shown in Fig.~\ref{fig:xrr}. The XRR measurements were carried out on SiO$_2$ and additionally on sapphire substrates. The qualitative intensity behaviour was highly reproducible. All data are from one series on SiO$_2$.

Four of the five hybrid structures clearly indicate Kiessig fringes. The most pronounced fringes can be detected for the 5~at.$\%$ and 10~at.$\%$ Al doped compounds: One can distinguish between two different oscillations hinting at the two layers, P3HT and nc-ZnO:Al, respectively, and a surprisingly small roughness at their interface. The spectra of the undoped and 0.8~at.$\%$ Al doped samples comprise fringes, as well. The spectrum of the 2~at.$\%$ Al doped ZnO/P3HT does not show pronounced interference features indicating a high level of interpenetration of the organic and the inorganic component accompanied by a high effective interface roughness.

To obtain $quantitative$ information about the thickness of P3HT and ZnO and about the interface roughness it is essential to perform simulations and fitting of the experimental data. 
This allows a direct determination of the interface roughness related to the infiltration of P3HT. The theoretical reflectivity was calculated by an optical matrix method, with a refinement procedure minimizing the difference between the calculated and the measured XRR curve: the program MOTOFIT~\cite{nelson2006} based on the Paratt formalism\cite{paratt}. 
The optimization of the parameters was done by minimizing the chi square in logarithmic form. Here, $\chi^2$ values around 5~$\%$ could be obtained
The real part of the scattering length density (SLD) which is directly related to the mass density and the atomic number ranges between 30.0~$\cdot$10$^{-5}$\AA$^{-2}$ and 35.5~$\cdot$10$^{-5}$\AA$^{-2}$ for ZnO and between 3.0~$\cdot$10$^{-5}$\AA$^{-2}$ and 9.3~$\cdot$10$^{-5}$\AA$^{-2}$ for P3HT. The layer thickness of ZnO:Al ranges between 12 and 23~nm, for P3HT between 19~nm and 26~nm. The extracted data including the electron density, derived from $\rho$=SLD/$r_e$, with $r_e$ the classical electron radius, are shown in Table~\ref{table}.

\begin{table*}
\caption{Electron density $\rho$, thickness d and rms roughness r of the ZnO:Al layer resulting from fits of the XRR spectra shown in Fig.~\ref{fig:xrr}.}
\begin{center}
\begin{tabular}{|c||c|c|c|c|c|}
\hline  & 0 at.$\%$ & 0.8 at.$\%$ & 2 at.$\%$ & 5 at.$\%$ & 10 at.$\%$ 
\\\hline\hline $\rho_{ZnO:Al}$ (e$^{-}$\AA$^{-3}$) & 1.1 & 1.1 & 1.3 & 1.1 & 1.2 
\\\hline $\rho_{P3HT}$ (e$^{-}$\AA$^{-3}$) & 0.1 & 0.2 & 0.1 & 0.2 & 0.3
\\\hline $\rho_{interlayer}$ (e$^{-}$\AA$^{-3}$) & 0.1 & 0.4 & 0.2 & 0.9 & 1.0
\\\hline d$_{ZnO:Al}$ (nm) &15 & 16 & 12 & 17 & 23 
\\\hline d$_{P3HT}$ (nm) & 20 & 19 & 23 & 26 & 25 
\\\hline d$_{interlayer}$ (nm) & 4 & 5 & 19 & 1 & 2
\\\hline r$_{ZnO:Al}$ (nm) & 2 & 1 & 1 & 0 & 1 
\\\hline \end{tabular}
\end{center}
\label{table}
\end{table*}

The fits based on the Paratt formalism are shown in Fig~\ref{fig:xrr}a. In order to obtain appropriate fits, it was necessary to assume intermixing of the P3HT and the ZnO:Al phases. A thin intermixing layer between 1 and 5~nm is estimated for 0, 0.8, 5 and 10~at.$\%$ Al doping and a very high value of 18.6~nm for 2~at.$\%$ as shown in Fig~\ref{fig:xrr}(b).

\section{Discussion}

Al doping of ZnO results in modifications of the electron transport, which will be discussed first. Remarkably it also influences the hole mobility of the subsequently applied P3HT. The reasons for the latter will be discussed in the second part of this section.
 
The mobility of the inorganic ZnO decreases over an order of magnitude with increasing Al doping and it remains nearly unaffected by the infiltration of P3HT. Unlike in the sintered system~\cite{hammer2008}, we observe two mobility plateaus (see  Fig.~\ref{fig:vt_mun}). The mobility stays at a nearly constant level at 2 to 3$\cdot$10$^{-3}$~cm$^2$V$^{-1}$s$^{-1}$ for the undoped, the 0.8~at.$\%$ and the 2~at.$\%$ Al doped ZnO. Also no significant change could be observed for 5 to 10~at.$\%$ Al doping. We attribute this effect mainly to scattering of electrons: We assume, the transport in $unsintered$ ZnO:Al is preferentially limited by charge carrier scattering effects due to the low level of crystallinity in the material, additional defects due to doping do not have a drastic effect on the mobility. It is worth mentioning, that the investigated layers are X-ray amorphous. The values for $\sigma$ and and the corresponding charge carrier density $n$ of around 10$^{-4}~\Omega^{-1}$m$^{-1}$ and 10$^{17}$~cm$^{-3}$ are rather low for ZnO:Al due to the omitted sintering step and, therefore, the low level of crystallinity. Also, errors due to the contact resistance can not be neglected. However, we expect this effect to be small due to the long channel length, i.e. the transport within the semiconductor limits the current between source and drain.

The hole mobility was found to depend on the Al doping level of ZnO. The origin of this behavior will be discussed in the following:

(i) We assume the hole transport to occur solely in P3HT, e.g. we modify the surface conditions of the undoped and doped zinc oxide layer and observe the influence on the morphology of the polymer. The hole mobility, again, is influenced by the morphology of P3HT, i.e. determined by the order of $\pi$-$\pi$-stacking and also by the orientation to the substrate because of the anisotropy of charge transport~\cite{kline2006}, and by the nanostructure of infiltration. Absorption spectra (not shown) revealed interchain absorption and a similar vibronic structure and therefor a similar degree of short range order of the polymer chains on top of the variously Al doped ZnO. %We note that optical absorption is a bulk sensitive technique, while the FET mobility is an interface property. However, in case of very thin layers and of layers with crystalline fraction induced by the underlying surface, the optical absorption spectra yields to (indirect) information on the interface topography. 

(ii) The porosity of ZnO determines the ability to infiltrate P3HT. 
The hole mobility sensitively depends on the roughness of the Al doped ZnO layer. XRR spectra exhibit a strong interconnection of the organic/inorganic interface layer and the doping level. The sample with 2~at.$\%$ Al doping shows a high level of intermixing, i.e. a value of around 18.6~nm (see Fig.~\ref{fig:xrr}) and also the highest hole mobility (see Fig.~\ref{fig:vt_mun}) in the set of doping levels. The effective thickness of the dielectric alone can not explain the origin of the hole mobility dependence on the doping level of ZnO. Nevertheless, the highest roughness value coincides with the highest hole mobility value.

(iii) The polarizability of the dielectric, i.e. ZnO:Al in the case of hole accumulation, influences the mobility of holes accumulated at the organic/inorganic interface. 
Fr\"ohlich polarons, reported for organic single crystals with different dielectrics by Hulea~$et~al.$\cite{hulea2006}, where the mobility depends on the value of the dielectric constant of the oxide, cannot describe $\mu_p~vs.~doping~level$, due to the necessity of a largely varying dielectric constant with doping level. A $\mu\sim1/\epsilon$ behaviour was found. A change in $\epsilon$ of almost two orders of magnitude would be mandatory in our case, e.g. from 8 to 0.08. In the case of ZnO, we can estimate a change in $\epsilon$ from $\sim$8 to $\sim$11~\cite{young1973}, if Al$_2$O$_3$ is present in a high percentage at the interface at high Al doping levels. Hence, Fr\"ohlich polarons can be excluded from having a significant influence on the dependence of the hole mobility on the Al doping level.
On the other hand, the mobility in disordered organic FETs was found to be diminished due to the systematic broadening of the Gaussian density of states with dipolar disorder at the semiconductor/dielectric interface due to the value of $\epsilon$~\cite{veres2003}. This scenario can also be excluded from having a drastic influence.

(iv) Defects or charged traps at the organic/inorganic interfaces generated by doping can positively or negatively influence the accumulated density of free holes.
In order to reach ambipolar behavior, a balanced intrinsic carrier density of both types of charge carriers is mandatory~\cite{pal2008}. If, e.g. the density of free electrons outbalances that of the holes, a higher negative gate voltage is needed in order achieve hole accumulation.
The threshold voltage for electron transport in bare ZnO:Al transistors has been investigated (Fig.~\ref{fig:sigmanVt}(a)). It ranges between 15.5~V and 33.4~V for the undoped and the 0.8~at.$\%$ Al doped ZnO and for the 5~at.$\%$ doped system, respectively, and reflects the doping dependence of $n$. 
The threshold voltage for holes (Fig.~\ref{fig:vt_mun}), which qualitatively indicates the density of trap states, nearly parallels the hole mobility. The trap scenario is affirmed by the fact that a low threshold voltage for electrons results in a high $V_{th}$ for holes after application of P3HT. The free electron density $n$ reaches a maximum in the 0.8~at.$\%$ and also in the 10~at.$\%$ Al doped samples. In both cases we observe low hole mobilities, namely 2.0$\cdot$10$^{-5}~$cm$^2$/Vs and 2.7$\cdot$10$^{-6}~$cm$^2$/Vs. A lower mobility value for the highest doping level is attributed to additional traps. The lowest charge carrier density was estimated to be 1.0$\cdot$10$^{16}$~cm$^{-3}$ and was observed for undoped ZnO, although 2~at.$\%$ Al doping with 1.6$\cdot$10$^{16}$~cm$^{-3}$ and 5~at.$\%$ with 1.8$\cdot$10$^{16}$~cm$^{-3}$ do not show a significant difference. If only the free electron density was responsible for the hole mobility dependence on the doping level of ZnO, the highest hole mobility should be observed in three cases: the undoped, the 2 and 5~at.$\%$ Al doped system. This is clearly not the case. 

Here, an interplay of different parameters, mainly the P3HT/ZnO:Al intermixing (ii) $and$ the free electron density in ZnO:Al (iv), causes the maximum hole mobility at 2~at.$\%$ Al doping. Interestingly, a minimum in $\sigma$, i.e., minimum in $\mu\cdot n$ can be found for the 5~at.$\%$ Al doped sample, where we indeed observe $balanced$ electron and hole mobilities after application of P3HT. Nevertheless, the $highest$ hole mobility was observed for 2~at.$\%$ Al doping. The Debye screening length in ZnO:Al at 0, 2, and 5~at .$\%$ Al doping, i.e. at the moderate electron densities of about 10$^{16}$~cm$^{-3}$,  amounts to $\sim$20~nm. This extends over the ZnO:Al thickness and hence, the free electrons will be depleted to the P3HT/ZnO:Al interface. In the case of the 2~at.$\%$ Al doped sample, where we observe the highest degree of intermixing, we expect local field enhancement at the P3HT/oxide interface and therefore better injection and accumulation of holes.

\section{Conclusion}

We were able to demonstrate ambipolar charge transport in a hybrid organic/inorganic transistor consisting of the polymer P3HT and the sol--gel synthesized undoped and Al doped ZnO. We observe the influence of doping of ZnO on its electron mobility and also on the hole mobility in P3HT. The maximum for the conductivity in unsintered ZnO has been reached for 0.8~at.$\%$ Al doping and also various Al doping levels strongly affect the density of free electrons. The electron mobility decreases with increasing doping level and stays nearly unaffected by the infiltration of the organic component. Remarkably, the hole mobility has been found to be dependent on the Al doping level of ZnO. The hole mobility reaches its maximum at the doping level of 2~at.$\%$. By discussing several possibilities, we attribute this effect to an enhanced intermixing of ZnO:Al and P3HT, derived via X-ray reflectivity data, which is necessary for the infiltration of P3HT and to the charge carrier density variation with Al contents. To summarize, the intrinsically free electrons and the morphology, i.e. intermixing, can be controlled via Al doping of ZnO and, hence, a balanced electron and hole mobility can be adjusted in P3HT/ZnO:Al hybrids. This is the fundamental precondition for the implementation of hybrid field effect transistors in organic electronics.

\section{Acknowledgements}
We acknowledge the BMBF (project GREKOS, 03SF0356A and 03SF0356B) and the EU (project Dephotex, NMP-SE-214458 and -214459) for financial support. C.D. gratefully acknowledges the support of the Bavarian Academy of Sciences and Humanities. V.D.Õs work at the ZAE Bayern is financed by the Bavarian Ministry of Economic Affairs, Infrastructure, Transport and Technology.

\appendix
\section{Electrical characteristics of ZnO:Al and ZnO:Al/P3HT hybrids in the linear regime}

\begin{figure} 
   \centering
   \includegraphics[width=8cm]{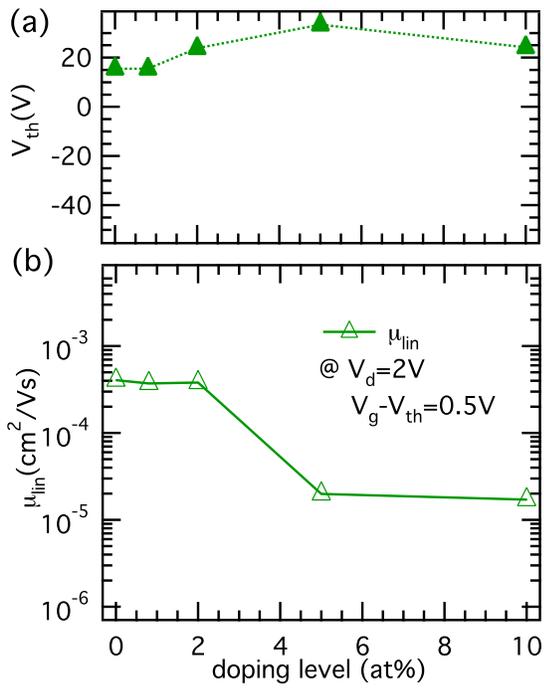} 
   \caption{(Color online) ZnO:Al characteristics, used for the evaluation of the free charge carrier density  $n$=$\sigma/\mu_e$ in bare ZnO:Al layers}
   \label{fig:znopuremu}
\end{figure}

\begin{figure} 
   \centering
   \includegraphics[width=8cm]{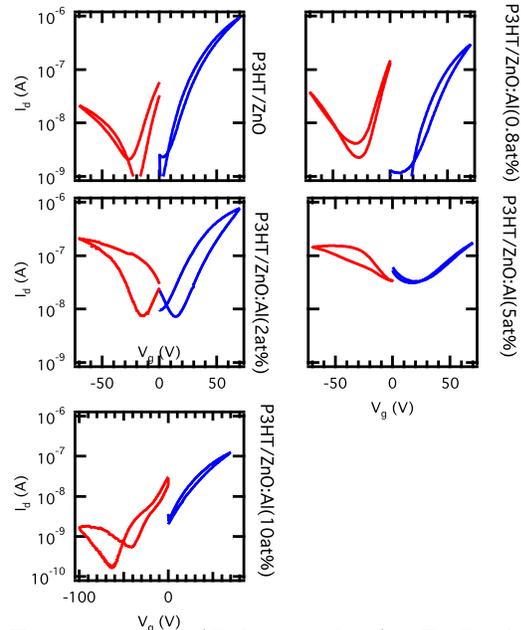} 
   \caption{(Color online) ZnO:Al/P3HT-Transistor: Linear transfer characteristics.}
   \label{fig:transferlin}
\end{figure}

Additionally to the data presented in Fig.~\ref{fig:sigmanVt} and Fig.~\ref{fig:transfersat} we present here the data in the linear regime for clarity. See Fig.~\ref{fig:znopuremu} and Fig.~\ref{fig:transferlin}. 

\section{Absorption}

\begin{figure} 
   \centering
   \includegraphics[width=8cm]{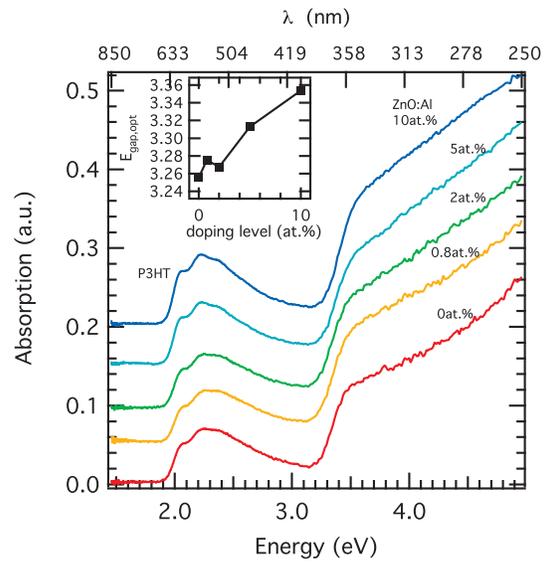} 
   \caption{(Color online) nc-ZnO:Al/P3HT absorption}
   \label{fig:absorption}
\end{figure}

The absorption spectra of the material system on sapphire were recorded using a PerkinElmer Lambda950.
We investigated absorption spectra in order to monitor the interchain orientation relative to the next neighbors of the infiltrated P3HT and to clarify the effect of doping on the absorption edge of ZnO:Al. 
The absorption spectra are shown in Fig. \ref{fig:absorption}.
We observe absorption by ZnO in the UV region at wavelength lower than 380~nm, in good agreement with literature~\cite{lin2008}. Using ($\alpha h\nu)^2 =A(h\nu-E_g)$ \cite{davismott1970}, where $A$ is a function of the refractive index of the material, the reduced mass and the speed of light in vacuum, $h$ is the PlanckÕs constant, $\nu$ is the frequency of the radiation and $E_g$ is the energy band gap. From a plot of $(\alpha h\nu)^2$ versus the photon energy an optical band gap was obtained by applying a linear fit and an extrapolation of $E_g$ to $(\alpha h\nu)^2$=0~eVm$^{-1}$. The undoped material exhibits an optical band gap of 3.26~eV (ca. 381~nm) and a shift of 100~meV to higher energies for the highest Al doping level was observed.

The polymer P3HT absorbs in the visible between $\sim$1.9~eV and $\sim$3~eV. It has been reported that enhanced ordering of this polymer gives rise to an increased absorption at  $\sim$2.09~eV~\cite{brown2003, chirvase2004}. The absorption spectra in Fig.~\ref{fig:absorption} indicate high ordering of P3HT for all doping levels.

\bibliographystyle{elsarticle-num}

\end{document}